# Experimental demonstration of non-adjacent band topology connecting multiple nodal links

Dongyang Wang[1], Biao Yang[1,2], Mudi Wang[1], Ruo-Yang Zhang[1], Xiao Li[1], Z. Q. Zhang[1], Shuang Zhang[3,4]*, C. T. Chan[1]*

1. Department of Physics and Center for Metamaterials Research, Hong Kong University of Science and Technology; Hong Kong, China.
2. College of Advanced Interdisciplinary Studies, National University of Defense Technology; Changsha 410073, China.
3. Department of Physics, The University of Hong Kong; Hong Kong, China.
4. Department of Electrical & Electronic Engineering, The University of Hong Kong; Hong Kong, China.
*Correspondence to: shuzhang@hku.hk; phchan@ust.hk

**Abstract**

Nodal links are special configurations of band degeneracies in the momentum space, where nodal line branches encircle each other. In *PT* symmetric systems, nodal lines can be topologically characterized using the eigenvector frame rotations along an encircling loop and the linking structure can be described with non-Abelian frame charges interacting among adjacent bands. In this paper, we present a photonic multiple nodal links system, where non-adjacent band topology is proposed to characterize the hidden relation between nodal lines from non-adjacent band pairs. Through an orthogonal nodal chain, the nodal line from the lower two bands predicts the existence of nodal lines formed between the higher bands. We designed and fabricated a metamaterial, with which the multiple nodal links and non-adjacent band topology are experimentally demonstrated.

**Introduction**

Topological photonics has emerged as a rapidly growing field[1-3]. Band touching or crossings in gapless topological materials can appear in momentum space as diverse forms, such as the Weyl points carrying nonzero chiral charges[4-9], and the line nodes accompanied with quantized Berry phase[10, 11]. The nodal lines in momentum space are typically protected by the mirror or *PT* symmetries[10-12], and can appear as straight lines, rings, chains, and links[13-27]. While topological characters are usually considered one band at a time resulting in integer topological invariants, there are recent efforts that consider simultaneously the properties of a group of adjacent bands (multiband approach) resulting in topological invariants that are elements of non-Abelian groups[28, 29]. Different from conventional topological band theory that uses the geometrical phase evolution of the eigenfunctions in a single band to define bulk topological invariants (e.g., Chern number), the multiband approach considers the frame rotation of a set of orthogonal eigenvectors (three or more) in Hermitian *PT*-symmetric systems. Since frame rotations are described by matrix-like entities, the multiband topological invariants are not integers, but are generalized quaternions, and their multiplications are generally non-commutative making the fundamental group non-Abelian. Such a non-Abelian approach actually predates modern topological band theory and can be traced to earlier work in classifying line defects in biaxial nematic liquid crystals [30-33]. The multiband frame rotation charges have been utilized to characterize the nodal line semimetals in the presence of *PT* (or $C_2T$) symmetry[26-29, 34-50], which provide new insight into the global structure of nodal lines in the momentum space.

In a *PT* symmetric system, the non-Abelian frame charges characterize the Hamiltonian by examining the real eigenvectors evolution along a closed homotopy ($\pi_1$) loop in the momentum space. These discrete rotations that bring a Hamiltonian back to itself form a non-Abelian generalized quaternion group[28, 36]. For example, the $\pi_1$ loop enclosed frame charge of $q = \pm g_i$ can be understood as the $\pm\pi$ rotations of a co-plane spanned by a pair of the eigenvectors ($i^{th}$ and $(i+1)^{th}$), which corresponds to a degeneracy node/line (between the $i^{th}$ and $(i+1)^{th}$ bands) encircled by the $\pi_1$ loop in the momentum space. More interestingly, 0 and $2\pi$ rotations become distinct and define the trivial and non-trivial charges that are noted as "+1" and "-1", respectively (Supplementary Information 1). The non-trivial charge of "-1" imposes extra constraints to the global nodal structure beyond the single band Berry phase description.

These non-Abelian frame charges can interact among adjacent bands through a common eigenvector, but the underlying topological connection between non-adjacent band nodal lines is not obvious. Here, by designing a photonic metamaterial that can be made and measured, we present a multiband system with multiple nodal links in momentum space. The Hermitian system possesses *PT* symmetry, and the nodal lines can be characterized using non-Abelian frame charges. By incorporating additional symmetry constraints, the topological connection between non-adjacent nodal lines is revealed. It is discovered that the configuration of nodal line formed between two lower bands explains the existence of additional nodal lines in the higher two bands via the interaction with nodal structure from intermediate bands, which established the underlying non-adjacent band topology. To illustrate our ideas, the photonic metamaterial is fabricated, and experimentally characterized at microwave frequencies, where the non-adjacent band topology predicted nodal lines have been experimentally observed.

**Multiple nodal links in the momentum space**

To exhibit multiple nodal links using metamaterials, more band degrees of freedom are needed beyond the fundamental propagating modes. This can be achieved using the split ring resonators (SRRs) as shown in Fig. 1(a). As commonly used in metamaterials, the SRRs possess both electric and magnetic resonances which provide longitudinal modes that can "cross" traverse modes to form nodal rings. The SRRs in Fig. 1(a) support an electric resonance driven by $E_y$ and a magnetic resonance driven by $H_z$. Here, the $E_x$-response can be neglected since the resonance frequency is much higher than the frequency regime of interest. The orientation of such two SRRs ensures the cancelation of bi-anisotropic couplings, i.e., $M_1$ and $M_2$ in Fig. 1(a) cancel out, which keeps the *PT* symmetry intact.

An effective medium description of the proposed SRR array results in the following constitutional parameters in the long wavelength limit as: $\varepsilon = [1, \varepsilon_{yy}, 1]$ and $\mu = [1, 1, \mu_{zz}]$, with $\varepsilon_{yy} = 1 + \frac{2l^2}{L(\omega_0^2 - \omega^2)}$ and $\mu_{zz} = 1 + \frac{1}{\omega_0^2 - \omega^2}\frac{2\omega^2 A^2}{L}$ (Supplementary Information 2), where we have set the non-resonant components to be unity for simplicity. Based on these effective parameters, we calculated the band dispersion as shown in Fig. 1(b). The resonances of $\varepsilon_{yy}$ and $\mu_{zz}$ introduce two longitudinal modes at the frequencies of $\omega_1$ and $\omega_2$ marked as stars. The longitudinal modes appear as flat bands along the $k_y$ and $k_z$ directions in Fig. 1(b), which intersect the transverse mode, forming into degeneracies marked with colored dots. The two frequencies of $\omega_1$ and $\omega_2$ correspond to the positions where the effective parameters $\varepsilon_{yy}$ and $\mu_{zz}$ go to zero and the effective medium becomes near-zero material[51], as confirmed by the dispersions of the constitutional parameters in Fig. 1(c).

We then retrieve the band degeneracy positions and show the results in the momentum space as Fig. 1(d), where multiple nodal links are found. In 3D momentum space, the intersections between the flat bands and the transverse mode form into two red nodal rings (formed between the 2$^{nd}$ and 3$^{rd}$ bands) in the perpendicular planes of $k_y = 0$ and $k_z = 0$, respectively. The two nodal rings touch at the $k_x$-axis and form as an orthogonal nodal chain in Fig. 1(d), where the intersecting modes coalesce along the $k_x$ direction as $\omega_a = \sqrt{k_x^2}$ and $\omega_b = \sqrt{k_x^2}/\sqrt{\varepsilon_{yy}}\sqrt{\mu_{zz}}$ (Supplementary Information 3). As can be noticed from Fig. 1(d), the blue nodal line (formed between the 1$^{st}$ and 2$^{nd}$ bands) threads through the orthogonal nodal chain in the momentum space. In addition, another pair of purple nodal lines (formed between the 3$^{rd}$ and 4$^{th}$ bands) also links with the orthogonal nodal chain in red. These nodal lines together form into the multiple nodal links in momentum space.

**Non-Abelian frame charges**

The multiple nodal links in the metamaterial can be characterized by non-Abelian frame charges. The Maxwell equations describing the metamaterial at low frequency can be encoded into an effective Hamiltonian as (Supplementary Information 4),

$$H = \begin{bmatrix} \frac{k_y^2}{1-2A^2} + k_z^2 + \omega_{px}^2 & \frac{k_x k_y}{-1+2A^2} & -k_x k_z & 0 & \frac{\sqrt{2}A\omega_0 k_y}{-1+2A^2} \\ \frac{k_x k_y}{-1+2A^2} & \frac{k_x^2}{1-2A^2} + k_z^2 + 2l^2 + \omega_{py}^2 & -k_y k_z & -\sqrt{2}\omega_0 & \frac{\sqrt{2}A\omega_0 k_x}{1-2A^2} \\ -k_x k_z & -k_y k_z & k_x^2 + k_y^2 + \omega_{pz}^2 & 0 & 0 \\ 0 & -\sqrt{2}\omega_0 & 0 & \omega_0^2 & 0 \\ \frac{\sqrt{2}A\omega_0 k_y}{-1+2A^2} & \frac{\sqrt{2}A\omega_0 k_x}{1-2A^2} & 0 & 0 & \frac{\omega_0^2}{1-2A^2} \end{bmatrix}$$

where the parameters $\omega_{px}$, $\omega_{py}$ and $\omega_{pz}$ are introduced as perturbations.

Using the eigenvectors of the effective Hamiltonian, non-Abelian frame charges are calculated (Supplementary Information 5) and presented as colored arrows in Fig. 1(d), where the color and the direction of arrow each represents the index "$i$" and "$\pm$" sign of the frame charge $q = \pm g_i$. We consider two $\pi_1$ loops indicated by green dotted lines in Fig. 1(d), with one in the horizontal and the other in the vertical plane, that are both encircling the $\Gamma$ point. Computation results show that the frame charges for the horizontal ($q_h$) and vertical loop ($q_v$) are both "-1". As $q_h = q_v = $ -1 simultaneously, we label the frame charge at $\Gamma$ as a "double -1" charge. The "double -1" charge is due to the braiding induced by electrostatic mode at photonic $\Gamma$ point, as has recently been proposed for biaxial media[41], only the material isotropy here is broken by resonance terms of $\varepsilon_{yy}$ and $\mu_{zz}$. These frame charges are essentially non-Abelian Berry phase and can be manifested as the rotation of eigenvectors or polarizations of photonic bands. We show the polarization rotations for the 1$^{st}$ and 2$^{nd}$ bands in Fig. 1(e) and (f) on the two mirror planes, where the $2\pi$ rotations around the $\Gamma$ point exhibit the "-1" element of the generalized quaternion group. This "double -1" charge imposes constraints on the multiple nodal links when the system is perturbed (Supplementary Information 6).

**Non-adjacent band topology**

One consequence of non-Abelian band topology is that the braiding of nodal lines, i.e., putting one nodal line in front of another, among adjacent bands leads to a sign change to the non-Abelian frame charge $q = \pm g_i$ of generalized quaternion group (indicated by flipping of arrow direction in the figures). Although the frame charges from distant bands are known to be Abelian [28, 29], little effort have been devoted to reveal the relation between non-adjacent nodal lines. We now proceed to reveal the non-adjacent band topology embedded in the multiple nodal links by taking symmetry constraints into account.

From the nodal line configuration shown in Fig. 1(d), we first isolate and show the intermediate red orthogonal nodal chain in Fig. 2(a), where an arrow configuration is assigned representing the non-Abelian frame charges $q = \pm g_2$ (characterizing the degeneracy between the 2$^{nd}$ and 3$^{rd}$ bands). The blue (1$^{st}$/2$^{nd}$ band degeneracy) and purple (3$^{rd}$/4$^{th}$ band degeneracy) nodal lines in Fig. 1(d) all link with the orthogonal chain, resulting in the braiding of nodal lines in the momentum space. We now take into consideration the blue nodal line as in Fig. 2(b), and we present the new arrow configuration after taking account only the braiding with the blue nodal line. We notice that the blue nodal line threads through the vertical nodal ring, and the braiding changes the direction of the arrows on the vertical nodal ring that are flowing to the left chain point, where the viewpoint (or base point for homotopy loops) can be selected as the "eye-symbol" indicated. It can then be found from Fig. 2(b) that the left and right chain points have different arrow configurations after the braiding. The difference in arrow configurations between the two chain points are, however, incompatible with the time-reversal symmetry (TRS), where topologically distinct frame charges ("+1" and "-1") are found for the two TRS-related $\pi_1$ loops shown in Fig. 2(b) as dashed circles. Note that the $\pi_1$ loop encircling the left chain point exhibits a charge of "-1", while the $\pi_1$ loop encircling the right chain point forms a charge of "+1". As "-1" and "+1" belong to different classes of the generalized quaternion group, they cannot be related by TRS.

Therefore, an additional topological structure must emerge to avoid this contradiction. Given the nodal structure of the 1st to 3rd bands in Fig. 2(b), one can infer that extra nodal lines are required to exist between the 3rd and 4th bands. One possible nodal line configuration to resolve the contradiction is presented in Fig. 1(d), where the vertical nodal ring is braided a second time by the purple nodal line. However, other configurations of purple nodal lines are also possible, such as the case shown in Fig. 2(c), as long as they thread through the orthogonal nodal chain to counterbalance the influence of the blue nodal line. We have hence shown that the purple nodal lines between the 3rd and 4th bands are topologically determined by the blue nodal line between the 1st and 2nd bands via interaction with the red orthogonal nodal chain from intermediate bands. These features exemplify the non-adjacent band topology that incorporates the fundamental constraints on the non-Abelian frame charges from additional symmetry requirements. Such constraints should be appliable to systems with more bands or carrying other forms of nodal structure.

**Metamaterial realization and experimental characterization**
We now build the metamaterial to realize the multiple nodal links system supporting non-adjacent band topology. The band structures of the metamaterial is numerically calculated (with CST Microwave studio) as shown in Fig. 3(a), where the resonators are arranged with periodicities of $p_x$ = 6 mm, $p_y$ = 4 mm, and $p_z$ = 3 mm as shown in the inset. Three mirror symmetries of $M_{x,y,z}$ are kept to protect the nodal line degeneracies indicated with colored dots in Fig. 3(a) (same color convention as nodal lines). Fig. 3(b) displays the BZ information highlighting the high symmetry lines. In Fig. 3(c), we show the retrieved nodal structure in the momentum space, where the multiple nodal links in the metamaterial agree well with the theoretical prediction from the effective medium model. One minor difference is in the location of the predicted nodal line between the 3rd and 4th bands, which, however, equivalently satisfies the non-adjacent band topology within the multiple nodal links as the case in Fig. 1(d) or Fig. 2(c) (Supplementary Information 7). The frame charges in the real metamaterial can be numerically checked through polarization rotations (Supplementary Information 8). Interestingly, the connection between the red and purple nodal lines takes the form of in-plane chain point in Fig. 3(c). The chain point between them leads to an apparent triple degeneracy (marked with a black star in band structures of Fig. 3(a) along Γ to Y), which is in fact an approximate accidental degeneracy and the exact degeneracy can occur on fine tuning of geometric parameters. However, the global linking between red and purple nodal lines is topologically protected. The triple degeneracy point adiabatically transforms to earring nodal links with equivalent topology when geometrical parameters change (Supplementary Information 9).

The proposed metamaterial is fabricated using printed circuit boards (PCBs) technology (with a substrate material of $\varepsilon_b \approx 2$), and further experimentally characterized to examine the proposed multiple nodal links. The electromagnetic field on sample surface is measured by a microwave near-field scanning system and the measured field are Fourier transformed to retrieve the momentums space information, including the projected bands and EFCs (Supplementary Information 10). We first study the fabricated sample at the x – z surface (80×10×80 units), where the vertical nodal ring in red can be experimentally observed. In Fig. 3(d), we show the calculated projected band dispersions along several $\theta$ directions ($\theta$ is defined in Fig. 3(c)). The crossed points on nodal ring along $\theta$ directions are manifested as the band degeneracy positions indicated with red dots in Fig. 3(d), where the red lines represent the bulk bands on $k_y = 0$ plane. The experimental measured results are shown in Fig. 3(e), which are in good agreement with the calculation results, and the predicted degeneracy points are indicated in red. In addition, the light cone positions are shown as the green lines, within which, the modes are leaky, and the measurement results show lower intensity.

In the first row of Fig. 3(f), we show the calculated equifrequency contours (EFCs) on the $k_y$ – $k_z$ plane for the 1$^{st}$ and 2$^{nd}$ bands at a few frequencies (Supplementary Information 10), and the experimentally measured results on the y – z sample surface (10×80×80units) are shown in the second row. The blue nodal line in the $k_x = 0$ plane is manifested as intersection points of bulk EFCs, which can be identified from the projected band EFCs in the measured results. We have thus observed the blue nodal line formed between the 1$^{st}$ and 2$^{nd}$ bands.

With another sample surface (80×80×5 units) shown in Fig. 4(a), we characterize the nodal structures in the $k_x$ – $k_y$ plane. The horizontal red nodal ring (formed between 2$^{nd}$ and 3$^{rd}$ bands) and the purple nodal lines (formed between 3$^{rd}$ and 4$^{th}$ bands) are shown in Fig. 4(b). To verify these two nodal lines, we cut the surface BZ at several lines corresponding to $k_x = [0.7, 0.8, 0.9]$ $\pi/p_x$, where the two nodal lines are separated and can be each identified. In Fig. 4(c), we calculated the band projection (black) and bulk bands for $k_z = 0$ (red) for such discrete $k_x$ values (scan along $k_y$). The nodal points on the red nodal ring are manifested as band crossings at lower frequencies (red dots), and the nodal points from purple nodal lines can be found at higher frequencies (purple dots). In Fig. 4(d), we show the experimentally measured band projections corresponding to the calculations, where the predicted band crossings can be identified. We thus have experimentally verified the predicted purple nodal line from non-adjacent band topology. Further surface modes can also be found from the x – y surface as presented in Supplementary Information 11.

In conclusion, we have revealed the non-adjacent band topology in the multiple nodal links, where nodal lines from distant band pairs are topologically related through intermediate nodal structures. The experimental observation of predicted nodal lines verified the non-adjacent band topology satisfying symmetry requirements that we proposed here. Our findings expanded the framework of non-Abelian frame charge and could motivate the exploration of further non-Abelian physics. The results establish metamaterials or photonic crystals as powerful and convenient platforms for exploring non-Abelian band topology, which may also inspire the design of new functioning topological devices.

**Acknowledgments**

This work is supported by Research Grants Council of Hong Kong through grants 16307821, AoE/P-502/20, 16310420, 16307621 and by KAUST20SC01.

**Figures**

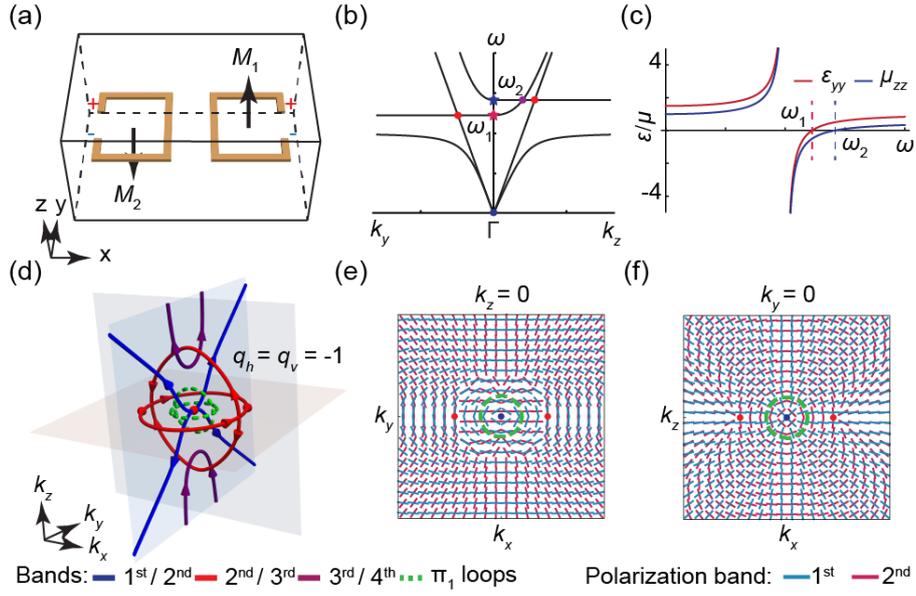

**Fig. 1. Multiple nodal links in the momentum space.** (a) Resonators for building multiple nodal links. The bi-anisotropic couplings in two rings cancel out. (b) Band dispersions along the $k_y$ and $k_z$ directions. Degeneracy points are colored in consistence with nodal lines. The adopted parameters are $l = 1$, $A = 0.5$, $\omega_0 = 2$, $\omega_{px,y,z} = 0$. (c) Dispersions of $\varepsilon_{yy}$ and $u_{zz}$, the frequencies at the zero values correspond to the flat bands in (b). (d) Nodal structures in momentum space, and multiple nodal links are formed. The two $\pi_1$ loops define the "double -1" charge at the $\Gamma$ point. (e-f) $2\pi$ eigenpolarization state rotations for the 1st and 2nd bands on the $k_z = 0$ and $k_y = 0$ planes, verifying the "double -1" charge.

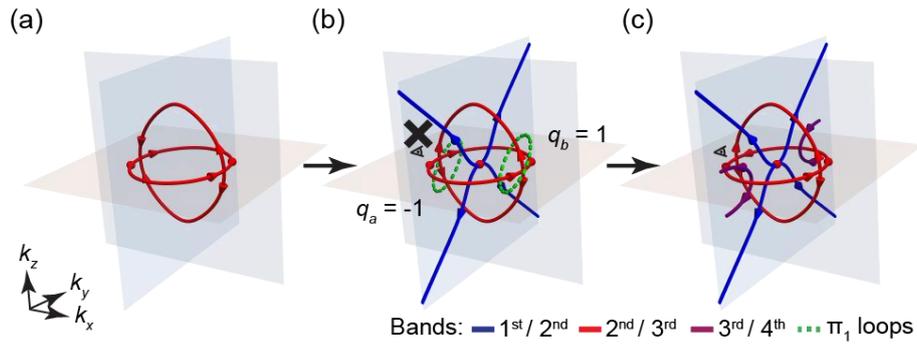

**Fig. 2. Non-adjacent band topology.** (a) Orthogonally chained nodal rings between the $2^{nd}$ and $3^{rd}$ bands in the momentum space. The initial arrows (representing the non-Abelian frame charges $q = \pm g_2$) are adopted in later analyzation. (b) The arrow configuration after considering the braiding with adjacent blue nodal line, with the viewpoint indicated with an "eye" symbol. Inconsistency is found for the two TRS-related $\pi_1$ loops. (c) Possible configuration of the predicted purple nodal lines, which counterbalances the braiding induced by the blue nodal line.

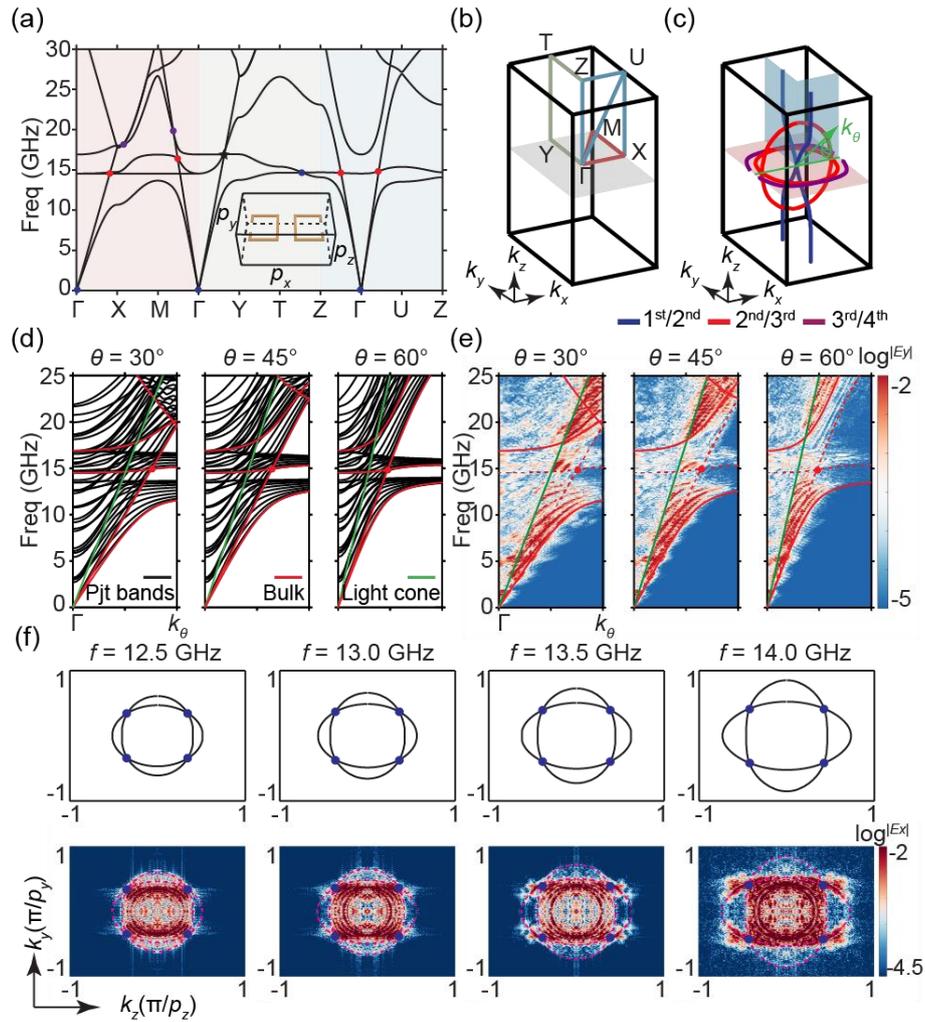

**Fig. 3. Photonic metamaterial realization of multiple nodal links.** (a) Band structures for the designed metamaterial (shown as inset, with background $\varepsilon_b = 2$). Resonator sizes are $L_x = 2$ mm, $L_y = 2.5$ mm, and gap $g = 1$ mm. (b) First BZ of the metamaterial. (c) Nodal structure in momentum space, with colours specifying the interacting bands. (d) Calculated band projection on the $k_x$ - $k_z$ plane along $k_\theta$-directions, with $\theta$ defined in (c). (e) Experimentally measured band projections along the $k_\theta$-directions, bulk bands are shown on top as red lines. (f) Calculated and measured EFCs in the $k_y - k_z$ plane. The marked points form into the blue nodal line in (c).

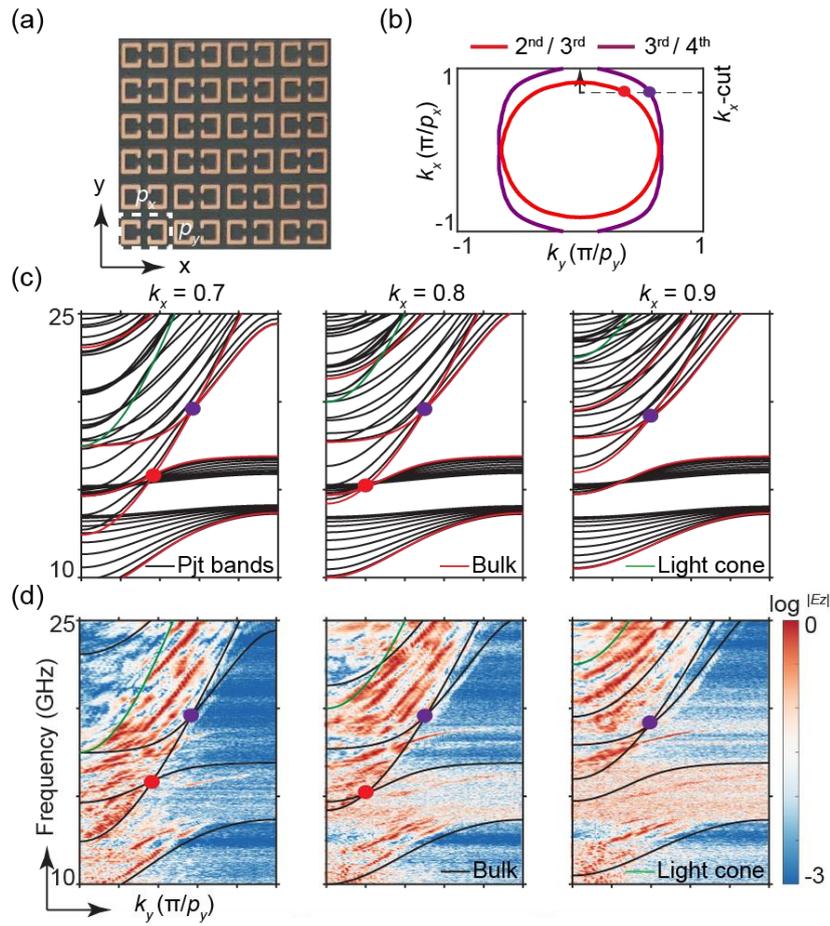

**Fig. 4. Experimental verification of the non-adjacent band topology.** (a) Photo of the fabricated metamaterial in the x – y plane. (b) Nodal structure in the $k_z = 0$ plane, the nodal ring from the 2nd and 3rd bands is shown in red, the predicted nodal lines from the 3rd and 4th bands are shown in purple. (c) Calculated band projections at fixed values of $k_x = [0.7, 0.8, 0.9]\,\pi/p_x$, where red nodal ring and purple nodal lines are manifested as colored points. Bulk bands at $k_z = 0$ plane are shown as red lines. (d) Experimentally measured band projections, with bulk bands shown in black.